\documentclass{article}
\usepackage[utf8]{inputenc}

\title{Analysis of Complex System Development Based on Fuzzy Cognitive Mapping}
\author{Ismail Kayali\thanks{M.S.c Big Data Systems, Faculty Business and Management, National Research University Higher School of Economics. Moscow,Russia. E-mail: ikayali@edu.hse.ru}}
\date{Feb 05 2018}

\begin{document}

\maketitle

\begin{abstract}

This article represents one of the contemporary trends in the application of the latest methods of classification in business, where intense competition and the desire to expand drive this science to far-reaching prospects using the discusses months and the most recent classification and forecasting algorithms such as SVM, FFM, C4.5, which are used to build better business decision support models, including basic steps in data pre-processing such as Attributes using Information Gain Ratio and filling missing values with several algorithms:
\textbf{K-Means},\textbf{ K-Nearest Neighbor},\textbf{ Linear Regression},\textbf{ Neural Network(Back Propagation)}

\end{abstract}

\section{Introduction}
Engineering and contemporary science are based on the first-principle models of biological, social and physical systems.
A primary scientific model, such as Newton's laws or Maxwell's equations, begins in electromagnetism and then builds various applications in mechanical or electrical engineering. In this method, empirical data are used to validate the models of the first principle and to estimate some of the unknown or imprecise parameters directly. In many areas, the first basic principles are unknown, or the systems under study are complex to be mathematical. With the expanding use of computers, there is a large amount of data generated by these systems. In the absence of the first principle models, these available data are used to derive models for estimating Useful relationships between system mutants (e.g. unknown links to output links). The need to understand large, complex and informative databases is practically common to all fields of work, engineering, and science. In the world of work, customer and company data are a source of strength where the usefulness of extracting useful knowledge hidden in these data and acting on this knowledge is increasingly important in today's competitive world. The whole process is called the application of the computer-based approach, including new techniques to discover the knowledge of data from data mining. Data mining is a repetitive process where progress is known to be detected either by mechanical or manual means.
It is very useful in the exploratory analysis scenario where there are no preconceived ideas about what the output will be.
It is the search for new information and non-primitive value in big data.
It is a collaborative work between human and computer where the best results achieved by balancing the knowledge of specialists in describing the problems and objectives with the capabilities of the search for computers. Do you believe in accidents in life?
As an analyst, it is likely that your answer is negative or to convey that you will learn it from now.
For example, a simple event you can imagine is throwing a coin. It is a random event; No one can predict the face that will appear after the heart. That may be true, but the fact that no one can predict it does not mean that it is impossible as a principle. Such as the speed of the throwing, the rotation angle, the properties of the constituent materials, the mass distribution, wind velocity, and direction, then we will be very able to some effort and time to predict the results of throwing a coin. However, the physical formulas of this relationship are known.
Let us now move on to another example; But this time we can predict the case out, the cup will be broken if it falls from a high altitude to some floor. We know that we will get a broken cup in few seconds we analyze the factors during the fall of the cup.
How do we have the proven ability to do such a thing? We have not seen the cup that falls at these moments broken before now, and in the end, the physical relations that describe the refraction of the cup are unknown to most of us.
Of course, the trophy may remain intact by "luck" in individual cases, but this is unusual.
The failure to break the cup is not just a "mismatch, luck" but a follow-up to the laws of physics. For example, the energy of the collision will go to Earth.
Well, how can people know what will happen in some cases and what will happen in other cases?
The most common explanation is that the interpretation of some cases is "congruence" and the description of other situations as "non-conformance. But let us put the following assumptions:
-The vast majority of the processes we recognize in our environment are not the result of chance.
-The inherent defect in our inability to accurately characterize and extrapolate processes is absolute because we are unable to identify or measure the factors with the underlying effect or correlations between them.
In the event of the fall of the cup we were able to quickly identify the most important features such as the quality of the material and fall from the height and nature of the ground, and in a very short time guess the possibility of refraction of the cup in comparison with previous experiences, but we can’t do this with the coin, we can see the throw of the piece number created But we will never succeed in identifying the necessary factors quickly enough and extrapolating the results in accordance with random firing.
So, what happened in our heads when we predicted the refraction of the cup after the collision?
We measured the characteristics of this event or to say that we collected the data describing the fall of the trophy and then we came in comparison to the result very quickly.
Where comparisons are made with previous fall experiences of cups, cups and similar materials based on similarity measures, the two necessary:
First, we need data for past events available; secondly, we need to realize the similarity between the present and all previously defined data. 
Finally, we will be able to guess or predict by looking for the most similar event in the past and placing it like the scene of the current event.
The search for the closest event to the current event is a kind of search for the ideal and named examples and gives us the closest and most accurate results.
Not necessarily every fall will lead to the refraction of the cup, but most of us can get the right guesses.
The relatively correct predictability of the future will enable us to address potential problems and avoid them and make the wiser decisions of the principle that events are similar and remain to be considered.
All of these done using data mining methods, automated learning and statistics that mimic human learning and adds other ways of seeing what we do not see and analyzing it so quickly that we can visualize the future and take advantage of it in real time.
The basic definition of decision making is:
A study to identify and choose between options and alternatives depending on the specifics and values, a process to reduce any hesitation about the options available to achieve a scientific and reasonable choice because sometimes our findings may be unreliable due to lack of research, and sufficient knowledge of alternatives with a good amount of Knowledge reduces risk and uncertainty but does not eliminate it.

\section{Decision Support Systems}

Decision support systems are directly targeted at decision-making while business intelligence provides accurate and timely information that indirectly supports the decision.
Means used in business intelligence are the same as those used in decision support systems such as (data mining and forecasting analysis)
Software companies develop the development of decision support systems by undergraduate and graduate students Business Intelligence.
For example, Harvard's 90-year decision support system was designed to predict which departments would need to be opened in the coming years or to calculate the best price for a new product for a business intelligence company.
Business Intelligence Objectives The objectives of decision support systems are mostly analytical[1].

\subsection{Decisions Types}
\textbf{Whether Decisions:}
Where decisions are yes, no / either, or.
The person decides whether the decision is achievable or not. Part of the decision-making process examines the pros and cons, and if the answer is "no," then another alternative study will be taken.

\textbf{Which Decisions:}
These decisions are taken from a range of alternatives and to be compared with the most likely option based on a set of criteria.

\textbf{Contingent Decisions:}
Decisions that are already known or taken but set aside until conditions are approved and met.
Most people make conditional decisions but keep them until they have the opportunity to apply the decision such as time, price, availability, motivation, energy, etc.

\section{Types of Decision Modelling}

\subsection{Kenper Tregoe Model}

These models are a unique way to solve problems systematically. There are four steps to this type of model:
\\1. Identify and evaluate the situation.
\\2. Problem analysis.
\\3. Evaluation of the decision.
\\4. Analysis of potential problems.

\subsection{Decision Step Model}

Sometimes called the logical decision model or the 8-step decision, but it ensures many variations as the steps are done sequentially step by step.
The situation and problem are important factors in determining the type of decision model to be used[2].

\subsection{Six Thinking Hats Model}

This model was developed to detect all angles and points in a complicated situation. The concept is six caps in multiple colors (green, yellow, blue, white, red, black) representing personal emotion/excitement.

\subsection{Carnegie Decision model}

Developed at Carnegie University, this model is a Satisficing strategy which is a combination of Satisfy and Sufficient.
Alternatives exist in research, and the least acceptable option is studied until a collective agreement is achieved[3].

\subsection{Iterative Decision Model}

This model is usually used in decisions involving techniques and steps that work increasingly and are tested from time to time.

\subsection{Vroom Yetton Model}

This model focuses on taking the most efficient and best decision, and also on the ways we reached this decision.
This sample contains a set of seven Yes / No questions, after which the decision criteria are calculated to take the appropriate case.

\subsection{Probability decision Model}

This model is a framework centered around two axes or two dimensions:
First, goal consent
Second, technical knowledge is the understanding of the relationship between cause and effect necessary to achieve the target.
Where a two-dimensional matrix is reached, the first dimension is a focus of consensus on the target, and the second dimension is specific to technical knowledge[4].

\section{Data Mining in Business}

Because of the large amounts of information are collected every day, and analysis of such data is a necessary need. Data mining can fit this need by providing means to discover knowledge from data, so data mining is a natural result of the information revolution.
The massive growth of information makes us live in the age of information. The information around us is many and scattered. We as analysts need to organize this information into structures to extract knowledge from it, which is the primary purpose of data analysis. That is, we live today in the age of knowledge and data mining. Optimize to derive this knowledge. There are several synonyms for Data Mining:
\\-Knowledge Mining from data.
\\-Knowledge Extraction.
\\-Data/pattern analysis.
\\-Data archaeology.
\\-Data dredging.

\section{Classification}
A classification is a form of data analysis that extracts models of important data. These models are called works that predict categorically (sporadically and unclassified) the label.
Most classification methods have been developed by researchers in the field of Machine Learning, Pattern Recognition and Statistics.
Most of these algorithms fully occupy memory and are ideal for small data size. The latest research in data mining is working on the development of classification and prediction algorithms that can handle large volumes of data.
The classification has a significant number of applications such as fraud detection, forecasting, production, medical diagnosis as well as marketing objectives.[6]

The first time it was described in the early 1950's and was not published until 1960 when the increase in computing power is possible, this algorithm is considered Lazy learner classifier.
The decision trees, SVM, and neural network are all classified as learning to learn. When a training set is given, it builds a classification before receiving Test data to organize it.
When Tuple training is provided, Lazy Learner is stored or performed by only a small fraction of the processing processes and is expected to be given a Test record.
When you see a log algorithm, It applies its classification based on its similarity with the saved training records.
Lazy Learner does less when he does a Training Tuple and does more work when he classifies or predicts digital. 
Lazy Learner is highly expensive and requires high storage techniques as it has
been studied very well to be applied in Parallel Hardware because it provides little insight or insight into the data structure.
Lazy Learner naturally supports growing learning. It can model high resolutions that have hyper polynomial forms that can’t be described as simple by other learning algorithms such as hyper rectangular forms shaped with decision trees.[4]

\section{Artificial Neural Network}

Artificial neural networks are one branch of artificial intelligence that mimic the neural networks in the learning of the machine. Neural networks have many important characteristics, including their ability to learn through complex models on new data models.[3]
The components of the artificial neural network and how to process information are the steps:
\\1-Treatment is done in simple processing elements called neurons.
\\2-Pass signals between neurons via interconnection lines.
\\3-Each weight line is accompanied by a certain weight, which is multiplied with
the input values to the neuron.
\\4- Apply on each neuron to activate the activation function to its input to determine the resulting output.

\section{Decision Trees}
Decision Trees is a form of nodes linked together by paths and under certain conditions that illustrate the necessary conclusions in the classification, the tree is built from the root to reach the branches so that they are connected to conditions that enable us to reach the solution and describe the problem more easily And illustrate the complex cases that need descriptive analysis, and the most famous decision tree algorithms are CART, C4.5, ID3.
At the beginning of the root is the owner of the largest profit and then the fork is based on the multiple values of that selected character up to one of the tree papers, which represents one of the items that account for the output of the classification process and thus the decision tree generated consists of two types of contract:
\\1- Decision nodes – represented by squares.
\\2-Chance nodes – represented by circles.
\\3-End nodes – represented by triangles.

Decision trees have three types of nodes and two branches.
Branches emerging from node resolution are decision-making branches; Each branch represents the available alternatives or the event path available at that node; the set of choices must be mutually exclusive (if one of the remaining options is selected) and must be aggregated (all available alternatives must be included / present in the group).

Each terminal node has a final value associated with it, called the resulting value. Each final value is calculated as a result of the following scenario:
The sequence of decisions and events on an individual chosen path that leads from the primary decision node to the specified final decision node.[7]
To determine the final value; This is done by using the method of assigning the values (weights or cash flow according to some references) to the decision branches and the event branches. The weights are then collected on the branches leading to the final node to determine the final value. In some problems, we need a model with more detailed values to determine the final values.

\section{Decision Trees Algorithms}

Most decision trees adopt the division strategy known as "divide and conquer." In this chapter, we will discuss one of these algorithms, the "C4.5" algorithm, an extension of the "ID3" algorithm after its disadvantages have been improved.

\subsection{ID3 algorithm}

In learning resolution trees, ID3 is an algorithm invented by ROSS Quinlan and used to generate the decision tree from the given data set.
Typically, ID3 is utilized in the learning machine in the field of natural language processing. Decision tree technology involves constructing a tree to model the classification process. Once the tree is built, it is applied to each line in the Tuple database and also applied to the results in another classification.
The ID3 algorithm is an information algorithm based on Information Entropy. Its
basic idea is that all samples are assigned to different classes according to various values of the attribute set state, and their essence is to determine the best classification attribute of the attribute sets.[6][9]
The algorithm selects Information Gain as a criterion for selection attributes, usually the attribute that has the highest information profit chosen as a division attribute for the current node, to make the interoperability of information divided into smaller subsets.
Branches can be constructed according to different values of attributes, and the previous process is repeatedly called for each branch to create other nodes and branches so that all the samples in one branch are subordinate to the same
attribute. The selection of the attributes of the division is dependent on the concepts of entropy and the gain of information.

\subsection{C4.5 algorithm}

The decision trees which created by the C4.5 algorithm can be used for classification, and therefore C4.5 refers to static classification.
The default criterion for selecting the Splitting Attribute attributes in the C4.5 algorithm is the Information Gain Ratio, rather than the information gain of the ID3 algorithm.
It is an effective method in the classification process and belongs to nodes and decision trees, which enable us to infer rules that are easily understood by the decision tree that we have. The tree is constructed based on the profitability values resulting from the measurement of the profit factor for each characteristic of the income group that will lead the process of building the tree based on one of the parameters used to measure the profit of the information.[8]

\section{Dataset Description}

The Dataset is a global competition that took place in October 2014 and extended to February 2015 through Kaggle (The Home of Data Science), a platform that offers competitions and business contracts for companies looking for analyzes and predictions of their data. The site contains several forecasting competitions, Movies and forecasting behavior across social media sites. We selected the contest from the Business Intelligence category.
Dataset belongs to AVAZU, an advertising company offering the following:
"Click-Through Rate CTR is an important criterion for evaluating ad performance, so the click expectation system will be substantial and widely used for funded research and real-time bidding."
"This competition is 11 days of information from AVAZU to build and test a prediction model. Can you find an algorithm that defeats the standard classification algorithms?"
Training group: 10 days in chronological order and 44 million records.
Test Group: A day of ads to test the forecast model with a million records.
The company did not give any other information and left the rest to the data scientists.
The data values are encrypted with the MD5 algorithm, which is a hash algorithm and therefore has no decryption, but some of the coding sites have been identified by the names of the countries where the ads are displayed. The company has
deleted the geographic area and modified the data to maintain confidentiality.

The large volume of data prevents the use of all dataset records, so it is supposed to be satisfied with one day of data as the data extends from 20 to 31 of the tenth month in 2014, but also the size of the data is very large, estimated one day with 4 million records so it lost the hours of 12 and 01 were taken from night 30 because the day closest to the test day has 178640 records to be studied.

\section{Using Data Mining in Business Intelligence}

It is necessary for entrepreneurs to earn a better understanding of the context of their systems such as customers, markets, resources, equipment, and competitors.
Without knowledge of data mining, many entrepreneurs can’t effectively analyze markets, compare customer feedback in similar products, discover the strength and weakness of competitors, retain their most valuable customers, and make smart business decisions.
It is clear that data mining is the essence of business intelligence, and analytical tools are dependent on multidimensional data mining.
An example illustrating the use of data mining in business intelligence:
Analysis of a company's loyalty model requires a classification model for potential customers leaving the company, for example, customers who may be left to the ISP and move to another.
This area of analysis is important for companies to gain new customers from their competitors. In this case, data scientists build new models every day to identify all the variables that may occur.
The model is then applied to all existing customers. According to this model, every customer who is likely to leave the company and has not received a special offer 30 days ago, the company offers a special offer ( Buying a mobile phone at a lower price). This offer increases the loyalty of the customer and reduces the possibility of moving to another company.
Based on the previous analysis of the loyalty model, the company manager has developed a strategy for campaigns to focus on increasing the loyalty of potential customers continually.

\section{Conclusion and Future Work}

There has been a significant similarity between the two concepts since the emergence of business intelligence.
Decision support systems are directly targeted at decision-making while business intelligence provides accurate and timely information that indirectly supports the decision.
Means used in business intelligence are the same as those used in decision support systems such as (data mining and forecasting analysis)
Software companies develop the development of decision support systems by undergraduate and graduate students Business Intelligence.
For example, Harvard's 90-year decision support system was designed to predict which departments would need to be opened in the coming years or to calculate the best price for a new product for a business intelligence company.
Business Intelligence Objectives The objectives of decision support systems are mostly analytical.

This Article discusses the most recently general classification and prediction algorithms like SVM, FFM, and C4.5.
Which are used to build better models to support decision making in Business Intelligence (BI) including essential steps for preprocessing of data. For example, the weighting of attributes using Information Gain Ratio and estimating and filling the missing values by using different algorithms such as:
K-Means, K-Nearest Neighbor, Linear Regression and Neural Networks (Back Propagation).

\end{document}